\documentclass[notitlepage,amssymb,amsmath,floatfix,nofootinbib,superscriptaddress]{revtex4-1}
\usepackage{epsfig}
\usepackage{bm}
\usepackage{amssymb}
\usepackage{amsmath}
\usepackage{color}
\usepackage{xcolor}
\usepackage{slashed}
\usepackage[colorlinks,linkcolor=blue,anchorcolor=black,citecolor=blue]{hyperref}
\usepackage{ulem}
\allowdisplaybreaks[4]

\begin{document}

\title{Helicity correlation of dihadron in current and target fragmentation regions of unpolarized SIDIS}

\author{Xue-Qi Xi}   
\affiliation{School of Science, Shandong Jianzhu University, Jinan, Shandong 250101, China}

\author{Kai-Bao Chen}
\email{chenkaibao19@sdjzu.edu.cn}
\affiliation{School of Science, Shandong Jianzhu University, Jinan, Shandong 250101, China}

\author{Xuan-Bo Tong}
\email{xutong@jyu.fi}
\affiliation{Department of Physics, University of Jyväskylä, P.O. Box 35, 40014 University of Jyväskylä, Finland}
\affiliation{Helsinki Institute of Physics, P.O. Box 64, 00014 University of Helsinki, Finland}

\author{Shu-Yi Wei}   
\email{shuyi@sdu.edu.cn}
\affiliation{Institute of Frontier and Interdisciplinary Science, Key Laboratory of Particle Physics and Particle Irradiation (MOE), Shandong University, Qingdao, Shandong 266237, China}

\author{Jing Wu}
\affiliation{School of Science, Shandong Jianzhu University, Jinan, Shandong 250101, China}

\begin{abstract}
We study the helicity correlation of two $\Lambda$ hyperons produced in unpolarized semi-inclusive deep inelastic scatterings (SIDIS), with one hyperon detected in the current fragmentation region and the other in the target fragmentation region. This observable provides direct access to the spin-dependent fragmentation function $G_{1Lq}$ and the spin-dependent fracture function $l_{1q}^L$ even in unpolarized lepton nucleon collisions. Utilizing the perturbative matching of the fracture function, we present numerical predictions for the helicity correlation, revealing significant variations with flavor and kinematic regions. This observable offers a unique way to investigate the spin-dependent hadronization mechanism across both the current and target fragmentation regions. It also provides new insights into the spin transfer effects in SIDIS processes.
\end{abstract}

\maketitle

\section{Introduction}

The intricacies of the hadron spin structure never cease to surprise. It offers deep insights into quantum chromodynamics~(QCD) from a unique perspective. Over the past few decades, a series of unexpected observations have kept the field vibrant. One notable example is the proton spin crisis~\cite{EuropeanMuon:1987isl, EuropeanMuon:1989yki}, first uncovered in polarized deep inelastic scattering experiments. These measurements demonstrated that quark spins contribute only a small fraction of the proton’s total spin, in stark contrast with the naive parton model prediction. Another remarkable case is the sizable transverse polarization of the $\Lambda$ hyperon observed in unpolarized proton-nucleus collisions~\cite{Bunce:1976yb, Heller:1983ia}, a phenomenon that remains puzzling within perturbative QCD and has motivated extensive theoretical studies. More recently, the Belle Collaboration has measured the transverse polarization of $\Lambda$ in $e^+e^-$ annihilation~\cite{Belle:2018ttu}. While the existence of such polarization is anticipated, the results have nevertheless triggered intense discussions on isospin symmetry in spin-dependent hadronization~\cite{DAlesio:2020wjq, Callos:2020qtu, Chen:2021hdn, DAlesio:2022brl, DAlesio:2023ozw, Li:2020oto}.

The hadronization of high energy partons is described by fragmentation functios~(FFs)~\cite{Metz:2016swz, Chen:2023kqw}. While significant progress has been made in understanding the unpolarized ones~\cite{Binnewies:1994ju, deFlorian:1997zj, Kniehl:2000fe, Kretzer:2000yf, Albino:2005me, Albino:2005mv, Kneesch:2007ey, deFlorian:2007aj, deFlorian:2007ekg, Hirai:2007cx, Albino:2008fy, Aidala:2010bn, deFlorian:2014xna, deFlorian:2017lwf, Anderle:2017cgl, Bertone:2017tyb, Bertone:2018ecm, Sato:2019yez, Khalek:2021gxf, Moffat:2021dji, AbdulKhalek:2022laj, Czakon:2022pyz,Gao:2024dbv,Li:2024etc,Gao:2025bko}, the spin-dependent FFs remain less explored due to the experimental and theoretical challenges. Yet, unraveling these spin-dependent FFs is essential, as they provide unique access to the nonperturbative dynamics of QCD, with hadron polarization serving as a powerful probe of the spin transfer and the spin–momentum correlation in hadronization. 

Already in the 1990s, the Large Electron-Positron Collider experiment~\cite{ALEPH:1996oew, OPAL:1997oem} measured the longitudinal polarization of $\Lambda$ hyperons. These experimental data can be used to investigate the longitudinal spin transfer in the hadronization of polarized partons \cite{deFlorian:1997zj,Boros:1998kc, Ma:1999wp, Boros:1999da, Filippone:2001ux,Wei:2013csa,Wei:2014pma,Chen:2016iey}. To understand the flavor dependence of the spin transfer, measurements in other polarized scattering processes have been called for. Various model estimations and phenomenological approaches have predicted sizable spin transfers for $\Lambda$ production in $e^+e^-$ annihilation, semi-inclusive deep inelastic scattering~(SIDIS), and hadron-hadron collisions for certain kinematic regions~\cite{Anselmino:1997ui, Boros:1998kc, Anselmino:1998jv, Ma:1998pd, Ma:1999gj, Ma:1999wp, Ma:1999hi, Anselmino:1999cg, Ma:2000uu, Ma:2000cg, Liu:2000fi, Liu:2001yt, Xu:2001hz, Zuo-tang:2002inz, Xu:2002hz, Ellis:2002zv, Xu:2003fq, Chi:2013hka}. 

In contrast, measurements in polarized SIDIS have found the $\Lambda$ polarization to be surprisingly small~\cite{HERMES:2006lro, COMPASS:2009nhs, Belostotski:2011zza, McEneaney:2022bsf}. The discrepancy between theoretical expectations and experimental data has thus again become a long-standing puzzle. A plausible resolution has been proposed recently in~\cite{Zhao:2024usu} by considering the possibility that hadron produced from the current fragmentation region (CFR) and the target fragmentation region (TFR) are not cleanly separated. Assuming a specific hadronization mechanism in the TFR, they find that a sizable TFR contribution would significantly dilute the $\Lambda$ polarization observed in inclusive measurements, and the prediction reconciles the SIDIS data. 

Contributions from the CFR and TFR are complementary in providing a complete description of hadron production in SIDIS~\cite{Boglione:2019nwk}. The CFR contribution is conventionally formulated within either the collinear factorization framework~\cite{Collins:1989gx, Wang:2019bvb, Daleo:2004pn, Kniehl:2004hf} or the transverse-momentum-dependent factorization framework~\cite{Collins:1981uk, Collins:2004nx, Ji:2004wu,Ji:2004xq, Collins:2011zzd, Boussarie:2023izj}, depending on the observables of interest. Within these frameworks, the nonperturbative dynamics of the initial and final states are encoded in the parton distribution functions~(PDFs) and FFs, respectively. However, such a separation is generally not applicable to SIDIS in the TFR. The description of hadron production in this region requires the introduction of fracture functions~\cite{Trentadue:1993ka, Berera:1995fj, Grazzini:1997ih}, which describe the joint distributions of the initial-state partons in the target and the observed hadrons originating from the target remnants.
In recent years, significant progress has been made in understanding hadron production in TFR and in elucidating the role of fracture functions~\cite{Anselmino:2011ss,Anselmino:2011bb,Anselmino:2011vkz,Chai:2019ykk,Chen:2021vby,Hatta:2022lzj,CLAS:2022sqt,Chen:2023wsi,Guo:2023uis, Tong:2023bus,Chen:2024brp,Chen:2024bpj,Hatta:2024vzv,Caucal:2025qjg}. In particular, Ref.~\cite{Chen:2024bpj} demonstrated that fracture functions can be interpreted as generating functions of the recently proposed nucleon energy correlators~\cite{Liu:2022wop} (see also~\cite{Cao:2023oef,Cao:2023qat,Liu:2023aqb,Li:2023gkh,Liu:2024kqt,Guo:2024jch,Guo:2024vpe,Mantysaari:2025mht,Gao:2025cwy,Huang:2025ljp}). These correlators have proven to be powerful tools for uncovering diverse QCD dynamics in the TFR, including small-$x$ phenomena~\cite{Liu:2023aqb,Mantysaari:2025mht}, spin physics~\cite{Li:2023gkh,Chen:2024bpj,Guo:2024jch,Guo:2024vpe,Mantysaari:2025mht,Huang:2025ljp,Gao:2025cwy,Song:2025bdj}, and even beyond-the-Standard-Model interactions~\cite{Huang:2025ljp}.

However, most studies in the TFR have focused on unpolarized hadron production, leaving the spin dependence of the final-state hadron in fracture functions largely unexplored. A systematic investigation of spin-dependent fracture functions is essential not only for understanding the hadronization mechanism and polarization effects intrinsic to the TFR, but also for establishing a more complete picture of hadronization across both the CFR and TFR.

The longitudinal spin transfer in CFR hadronization is characterized by the spin-dependent fragmentation function $G_{1Lq}$, where traditional studies usually rely on polarized collisions or electroweak processes~\cite{ALEPH:1996oew,OPAL:1997oem}. Recently, the spin correlation of dihadron~\cite{Chen:1994ar, Zhang:2023ugf, Li:2023qgj, Shao:2023bga, Chen:2024qvx, Yang:2024kjn, Huang:2024awn, Shi:2024gex,Shen:2024buh, Lv:2024uev} emerges as a powerful tool to probe spin-dependent FFs through unpolarized collisions, beyond their intrinsic connection to quantum entanglement phenomena~\cite{Barr:2024djo, Guo:2024jch, Cheng:2025cuv, vonKuk:2025kbv, Lin:2025eci, Qi:2025onf, Cao:2025qua}. It offers a complementary, and possibly more accessible, way for probing spin-dependent hadronization in a wide range of unpolarized collision processes, significantly enriching our toolkit. In this work, we propose the helicity correlation of hadrons in the current and target fragmentation regions as a novel probe to study the spin-dependent fragmentation functions and fracture functions in unpolarized SIDIS experiments.

The rest of this paper is organized as follows. In section~\ref{sec:framework}, we present the theoretical framework for calculating the dihadron production in SIDIS with one hadron in CFR and the other in TFR. The dihadron helicity correlation is defined. In section~\ref{sec:numerical}, we first calculate the perturbative matching of the fracture functions and then, based on the matching result, provide numerical predictions of the helicity correlation. A summary is given in section~\ref{sec:summary}.

\section{Dihadron production in SIDIS and the helicity correlation}
\label{sec:framework}

We consider the dihadron production in unpolarized SIDIS process,
\begin{align}
e(l)+N(P)\to e(l^\prime)+h_1(P_1,S_1)+h_2(P_2,S_2)+X~,
\end{align}
with the final state hadron $h_1$ produced in CFR and $h_2$ in TFR.
The momenta of the incident electron, the nucleon, the outgoing electron and the produced hadrons are denoted by $l$, $P$, $l^\prime$ and $P_1$, $P_2$, respectively.
At the leading order of quantum electrodynamics, one virtual photon is exchanged between the electron and the nucleon. The momentum of the virtual photon is given by $q=l-l^\prime$.
The spin states of the produced hadrons are indicated by $S_1$ and $S_2$.
To be specific, we will consider the production of a pair of $\Lambda$ hyperons in the following.

Throughout this paper, we use the light-cone coordinate system, in which a vector $a^\mu$ is expressed as $a^\mu = (a^+,a^-, \boldsymbol{a}_\perp) = \bigl((a^0+a^3)/{\sqrt{2}}, (a^0-a^3)/{\sqrt{2}}, a^1, a^2 \bigr)$. The set of Lorentz invariant variables used for this SIDIS process are defined by $Q^2 = -q^2$, $x_B = \frac{Q^2}{2 P\cdot q}$,  $y=\frac{ P\cdot q}{P\cdot l}$, $z_1 = \frac{P\cdot P_1}{P\cdot q}$, $\xi_2 = \frac{P_2 \cdot q}{P \cdot q}$~\cite{Graudenz:1994dq,Anselmino:2011ss,Boglione:2019nwk}. 
We work in the collinear frame of the virtual photon and the nucleon in which the nucleon moves along the $z$-direction and the virtual photon along the $-z$-direction.
In this frame, the momenta of the particles are given by
\begin{align} 
& P^\mu = ( P^+,0,\boldsymbol{0}_\perp)~,\\
& P_1^\mu = (P_1^+, P_1^-, \boldsymbol{P}_{1\perp})~, \\
& P_2^\mu = (P_2^+, P_2^-, \boldsymbol{P}_{2\perp})~, \\
& l^\mu = \Big(\frac{1-y}{y}x_B P^+,~ \frac{Q^2}{2x_B y P^+},~ \frac{Q\sqrt{1-y}}{y},~0\Big)~, \\
& q^\mu =\Big(-x_B P^+,~ \frac{Q^2}{2x_BP^+}, \boldsymbol{0}_\perp\Big)~.
\end{align}
For the case of $h_1$ produced in CFR and $h_2$ in TFR, we have $P_1^- \gg |\boldsymbol{P}_{1\perp}| \gg P_1^+$ and $P_2^+ \gg |\boldsymbol{P}_{2\perp}| \gg P_2^-$.

\begin{figure}[!htb]
\centering
\includegraphics[width=0.31\textwidth]{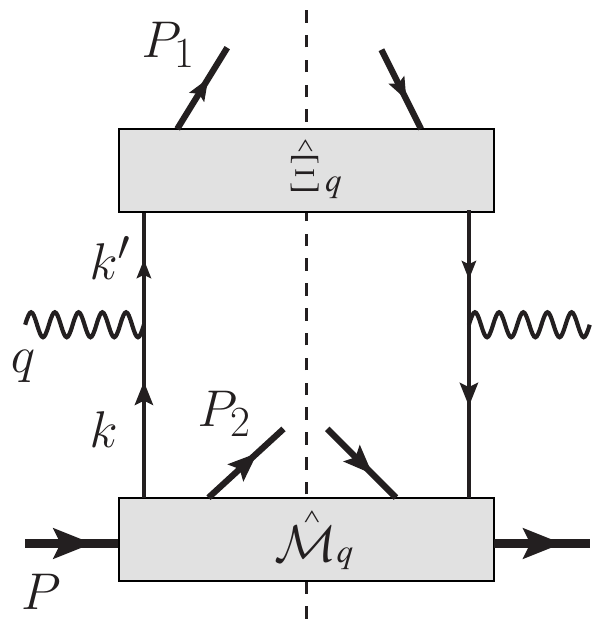}
\caption{Feynman diagram of the hadronic tensor at the tree level of perturbative QCD. Similar antiquark contribution is not shown.}
\label{fig:Wuv}
\end{figure}

Under one-photon exchange approximation, the differential cross section takes the form~\cite{Anselmino:2011bb}:
\begin{align}
\frac{d\sigma}{dx_B dy dz_1 d\xi_2 d^2\boldsymbol{P}_{1\perp} d^2\boldsymbol{P}_{2\perp} d\psi} = \frac{\alpha_{\rm em}^2 y}{4 z_1 Q^4} L_{\mu\nu}(l,l^\prime) W^{\mu\nu}(q,P,P_1,P_2)~, \label{eq:Xsection}
\end{align}
where $\alpha_{\rm em}$ is the fine structure constant, $\psi$ is the azimuthal angle of the outgoing lepton around the incident lepton beam.
The leptonic tensor $L^{\mu\nu}(l,l^\prime)$ is
\begin{align}
L^{\mu\nu}(l,l^\prime)=
2 (l^\mu l^{\prime\nu} + l^\nu l^{\prime\mu} - l\cdot l^\prime g^{\mu\nu})~.
\label{eq:Luv}
\end{align}
$W^{\mu\nu}(q,P,P_1,P_2)$ is the hadronic tensor.
At the tree level of perturbative QCD, it is represented by the Feynman diagram in Fig.~\ref{fig:Wuv}.
The hadronic tensor can be calculated as
\begin{align}
W^{\mu\nu} = \sum_q e_q^2 \int d^4k \int d^4 k^\prime \delta^4(k+q-k^\prime) {\rm Tr} \left[ \hat {\cal M}_q(k;P;P_2,S_2) \gamma^\mu \hat \Xi_q(k^\prime;P_1,S_1) \gamma^\nu \right] + \{q\leftrightarrow \bar q \}~,
\label{eq:Wuv-raw}
\end{align}
where $\{q\leftrightarrow \bar q\}$ denotes the contribution from the anti-quark.
The associated correlation matrix elements are defined as
\begin{align}
&\hat \Xi_{q,ij} (k^\prime;P_1,S_1) = \int \frac{d^4\zeta}{(2\pi)^4} e^{ik^\prime\cdot \zeta} \langle 0| \psi_i(\zeta) |P_1,S_1;X\rangle\langle X;P_1,S_1|\bar\psi_j(0)|0\rangle~, \\
&\hat {\cal M}_{q,ij} (k;P;P_2,S_2) = \frac{1}{2\xi_2 (2\pi)^3}\int \frac{d^4\eta}{(2\pi)^4} e^{ik\cdot \eta} \langle P|\bar \psi_j(0) |P_2,S_2;X\rangle\langle X;P_2,S_2|\psi_i(\eta)|P\rangle~.
\end{align}
The matrix $\hat \Xi_{q} (k^\prime;P_1,S_1)$ represents the fragmentation of a quark into a hadron.
The matrix $\hat {\cal M}_{q} (k;P;P_2,S_2)$ represents the parton distribution of the nucleon when a final-state hadron is detected in the remnants of the target nucleon. 
At the leading power of large $Q^2$, one can simplify the hadronic tensor by neglecting the small longitudinal components in the parton momenta.
Taking $k^\mu \approx (xP^+, 0, \boldsymbol{k}_T^\mu)$ and $k^{\prime\mu} \approx (0, P_1^-/z, \boldsymbol{k}_T^{\prime\mu})$ in the $\delta$-function of Eq.~(\ref{eq:Wuv-raw}), we can integrate over the $-$ component of $k^\mu$ and $+$ component of $k^{\prime\mu}$.
Further simplification can be made by considering the cross section integrated over the transverse momentum of $h_1$.
This can be done in the collinear frame of the nucleon and $h_1$.
In this frame, the virtual photon will have a transverse momentum $\boldsymbol{q}_T = -\boldsymbol{P}_{1\perp}/z_1$, which gives $d^2\boldsymbol{P}_{1\perp} = z_1^2 d^2 \boldsymbol{q}_T$.
Integration over $d^2\boldsymbol{P}_{1\perp}$ in the cross section of Eq.~(\ref{eq:Xsection}) will eliminate the transverse $\delta$-function in the hadronic tensor.
Then, we can further integrate over the parton transverse momenta $\boldsymbol{k}_T$ and $\boldsymbol{k}_T^\prime$ in the hadronic tensor and arrive at
\begin{align}
\frac{d\sigma}{dx_B dy dz_1 d\xi_2 d^2\boldsymbol{P}_{2\perp} d\psi} = \frac{\alpha_{\rm em}^2 y z_1}{4Q^4} L_{\mu\nu}(l,l^\prime) W^{\mu\nu}(x_B;z_1,S_1;\xi_2,P_{2\perp},S_2)~, \label{eq:Xsec}
\end{align}
with the hadronic tensor and correlation matrix elements given by
\begin{align}
& W^{\mu\nu}(x_B;z_1,S_1;\xi_2,P_{2\perp},S_2) = \sum_q e_q^2{\rm Tr} \Bigl[ \hat{\cal M}_q(x_B; \xi_2, P_{2\perp},S_2) \gamma^\mu \hat\Xi_q(z_1,S_1)  \gamma^\nu \Bigr] + \{q \leftrightarrow \bar q\}~,\label{eq:WuvTr} \\
& \hat \Xi_{q,ij}(z_1,S_1) = \int \frac{d\zeta^+}{2\pi} e^{iP_1^-\zeta^+/z_1} \langle 0| \psi_i(\zeta^+) |P_1,S_1;X\rangle\langle X;P_1,S_1|\bar\psi_j(0)|0\rangle~,\label{eq:XiFF} \\
& \hat {\cal M}_{q,ij}(x_B; \xi_2, P_{2\perp},S_2) = \int \frac{d\eta^-}{2\xi_2 (2\pi)^3} e^{ix_B P^+ \eta^-} \langle P|\bar \psi_j(0) |P_2,S_2;X\rangle\langle X;P_2,S_2|\psi_i(\eta^-)|P\rangle~.\label{eq:MFrF}
\end{align}
It is noted that there should be gauge links in the definition of $\hat \Xi_q$ and $\hat {\cal M}_q$ in order to keep gauge invariance.
We have suppressed the gauge links here for brevity.

The matrices of Eqs.~(\ref{eq:XiFF}) and (\ref{eq:MFrF}) can be decomposed in terms of Dirac $\Gamma$-matrices.
A complete decomposition including polarized hadron production case can be found, e.g., in~\cite{Chen:2016moq,Metz:2016swz} for $\hat\Xi_q$, and in~\cite{Anselmino:2011ss,Chen:2023wsi} for $\hat {\cal M}_q$.
Here, we only consider the longitudinal spin of the produced hadron.
Let $\lambda_1$ and $\lambda_2$ be the helicity of the produced hadron $h_1$ and $h_2$, under the constraints of parity, we have
\begin{align}
& z_1 \hat \Xi_{q,ij}(z_1,S_1) = (\gamma^+)_{ij} D_{1q}(z_1) + (\gamma_5\gamma^+)_{ij} \lambda_1 G_{1Lq}(z_1) + \cdots~, \label{eq:DecFF}\\
& \hat {\cal M}_{q,ij}(x_B; \xi_2, P_{2\perp},S_2) = \frac{1}{2}\Bigl[ (\gamma^-)_{ij} u_{1q}(x_B,\xi_2,P_{2\perp}) + (\gamma_5\gamma^-)_{ij} \lambda_2 l_{1q}^L(x_B,\xi_2,P_{2\perp}) \Bigr] + \cdots~. \label{eq:DecFrF}
\end{align}
Here, $D_{1q}$ is the ordinary unpolarized fragmentation function, and $G_{1Lq}$ is the polarized fragmentation function describing the longitudinal spin transfer from the quark to the hadron.
The functions $u_{1q}$ and $l_{1q}^L$ are called as extended fracture functions.
They depend not only on the longitudinal momentum fraction of the nucleon but also on the transverse momentum of the hadron produced in the TFR.
In the rest of this paper, we simply refer to them as fracture functions.
The fracture function $l_{1q}^L$ encodes the correlation of longitudinal spins between the initial struck quark and the final-state hadron observed in the TFR.
The $\cdots$ in Eq.~(\ref{eq:DecFrF}) denotes chiral-odd terms or contributions beyond twist-2 in the decomposition.
As shown in Eq.~(\ref{eq:DecFF}), no chiral-odd collinear fragmentation function exists at the leading twist or twist-2, so the chiral-odd fracture functions do not contribute.

Contracting the hadronic tensor in Eq.~(\ref{eq:WuvTr}) with the leptonic tensor in Eq.~(\ref{eq:Luv}), we obtain the differential cross section in terms of the associated FFs and fracture functions, given by
\begin{align}
\frac{d\sigma}{dx_B dy dz_1 d\xi_2 d^2\boldsymbol{P}_{2\perp} d\psi}
& = \frac{\alpha_{\rm em}^2}{yQ^2} \sum_q e_q^2 A(y) \bigl[ u_{1q}(x_B,\xi_2,P_{2\perp}) D_{1q}(z_1) + \lambda_1\lambda_2 l_{1q}^L(x_B,\xi_2,P_{2\perp}) G_{1Lq}(z_1) \bigr] + \{q\leftrightarrow \bar q\}~,\label{eq:Xsec-final}
\end{align}
where $A(y) = y^2-2y+2$.

We define the dihadron helicity correlation as~\cite{Zhang:2023ugf}
\begin{align}
{\cal C}_{LL} &\equiv \frac{d\sigma^{++} + d\sigma^{--} - d\sigma^{+-} - d\sigma^{-+}}{d\sigma^{++} + d\sigma^{--} + d\sigma^{+-} + d\sigma^{-+}}~,
\label{eq:CLL-def}
\end{align}
where $d\sigma^{\lambda_1 \lambda_2}$, with $\lambda_{1,2} = \pm1$, is the differential cross section for hadron production with specified helicity states.
Substituting Eq.~(\ref{eq:Xsec-final}) into Eq.~(\ref{eq:CLL-def}), we obtain
\begin{align}
{\cal C}_{LL}(x_B,\xi_2, P_{2\perp}, z_1) = \frac{\sum_q e_q^2 l_{1q}^L(x_B,\xi_2,P_{2\perp}) G_{1Lq}(z_1) + \{q\leftrightarrow \bar q\}}{\sum_q e_q^2 u_{1q}(x_B,\xi_2,P_{2\perp}) D_{1q}(z_1) + \{q\leftrightarrow \bar q\}}~.
\label{eq:CLL-in-FrF}
\end{align}
Eq.~(\ref{eq:CLL-in-FrF}) clearly shows that the helicity correlation is determined by the product of the polarized fragmentation function and the polarized fracture function, thereby encoding the correlation of longitudinal spin transfers from the initial struck parton to the final-state hadron observed in the CFR and the TFR, respectively. 
This correlation is accessible in unpolarized SIDIS experiments, offering a novel and practical probe for investigating spin-dependent hadronization mechanisms in both the CFR and TFR.

\section{Numerical predictions of the helicity correlation}
\label{sec:numerical}
To numerically estimate the dihadron helicity correlation derived in Eq.~(\ref{eq:CLL-in-FrF}), nonperturbative inputs of the relevant FFs and fracture functions are required. We consider the case, where both hadrons are $\Lambda$ hyperons, since the helicity of $\Lambda$ can be easily measured via its self-analyzing weak decay. Parameterizations and phenomenological fits of the $\Lambda$ FFs have long been available for both unpolarized and polarized cases~\cite{deFlorian:1997zj,Albino:2008fy}. However, no phenomenological parameterization currently exists for the fracture functions $u_{1q}$ and $l_{1q}^L$.

A simplified approach is to integrate over the transverse momentum of the hadron in the TFR, expressing the helicity correlation in terms of the corresponding integrated fracture functions~\cite{Anselmino:2011ss}:
\begin{align}
& U_{1q}(x_B,\xi_2) = \int d^2\boldsymbol{P}_{2\perp} u_{1q}(x_B,\xi_2,P_{2\perp})~, \\
& L_{1q}^L(x_B,\xi_2) = \int d^2\boldsymbol{P}_{2\perp} l_{1q}^L(x_B,\xi_2,P_{2\perp})~.
\end{align}
 Here, the transverse momentum ${P}_{2\perp}$ is integrated up to a scale that ensures the hadron $h_2$ remains within the nonperturbative region of the TFR. Most existing phenomenological analyses of integrated fracture functions have focused on the unpolarized leading-baryon production, i.e., $U_{1q}(x_B,\xi_2)$~\cite{deFlorian:1997wi,deFlorian:1998rj,Ceccopieri:2012rm,Shoeibi:2017lrl,Shoeibi:2017zha,Ceccopieri:2015kya}, whereas no phenomenological constraints are yet available for the polarized counterparts $L_{1q}^L(x_B,\xi_2)$. A recent study of spin transfer to the $\Lambda$ hyperon in SIDIS~\cite{Zhao:2024usu} suggests that, under the assumption of a specific fragmentation mechanism, the leading target-fragmentation channel predominantly produces unpolarized $\Lambda$ hyperons. Within this hypothesis, the polarized fracture function is negligible, leading to zero spin transfer in the TFR and, consequently, a vanishing dihadron helicity correlation.

While this picture may provide a reference in the nonperturbative small-${P}_{2\perp}$ regime, we extend the investigation to the intermediate transverse-momentum region, $\Lambda_{\text{QCD}} \ll P_{2\perp} \ll Q$. In this domain, perturbative contributions become relevant and are expected to yield non-zero helicity correlations. The transverse momentum of the observed hadron is then produced through perturbative parton splittings and radiations, allowing for the standard collinear factorization in terms of PDFs and FFs.

\begin{figure}[!htb]
\centering
\includegraphics[width=0.6\textwidth]{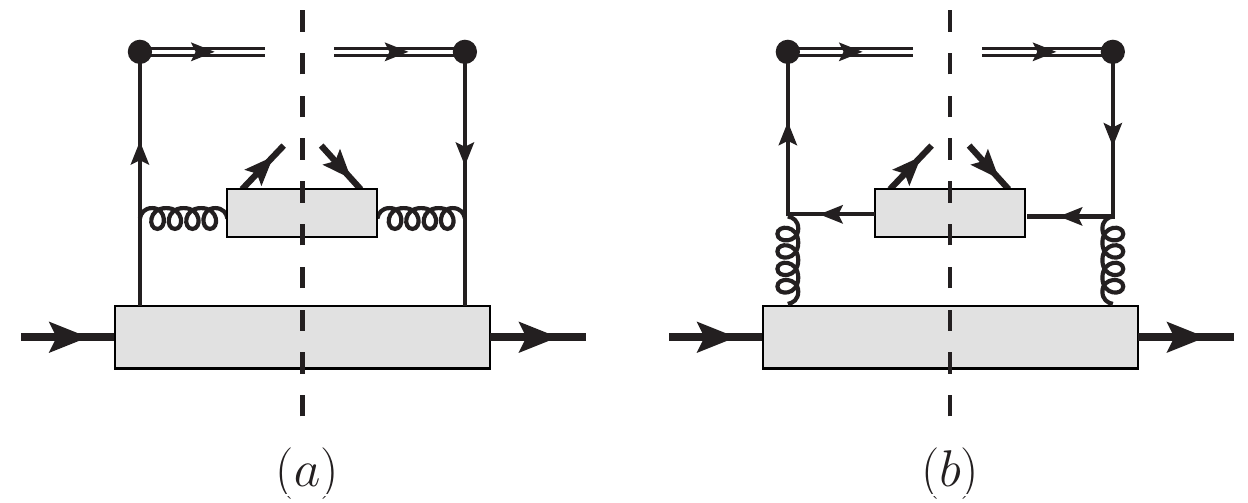}
\caption{Diagrams for the matching of quark fracture functions at the leading order of $\alpha_s$.}
\label{fig:Matching}
\end{figure} 

It has been demonstrated in~\cite{Chen:2021vby} that the unpolarized quark fracture function $u_{1q}$ for observing a $\Lambda$ hyperon in a proton target can be further factorized and perturbatively matched onto proton PDFs and $\Lambda$ FFs in the region $\Lambda_{\text{QCD}} \ll P_{2\perp} \ll Q$. This analysis can be straightforwardly extended to the polarized fracture function $l_{1q}^L$.
At leading power in the large-$P_{2\perp}$ expansion, both the unpolarized and polarized fracture functions match onto twist-2 distributions. As illustrated in Fig.~\ref{fig:Matching}, there are two perturbative diagrams contributing to the quark fracture functions in the light-cone gauge. In Fig.~\ref{fig:Matching}(a), the large transverse momentum $P_{2\perp}$ originates from the splitting of a quark from the unpolarized proton, followed by gluon fragmentation into a $\Lambda$ hyperon. In Fig.~\ref{fig:Matching}(b), the contribution arises from gluon splitting and subsequent fragmentation of an antiquark.
For completeness, we present the detailed matching calculations for both $u_{1q}$ and $l_{1q}^L$ in the Appendix.
The resulting matching formulas read:
\begin{align}
u_{1q}(x_B,\xi_2,P_{2\perp}) &= \frac{\alpha_s}{2\pi^2 \xi_2 \boldsymbol{P}_{2\perp}^2} \int_{x_B+\xi_2}^1 \frac{dx}{x}z \Biggl[ P_{gq}\left(\frac{\xi_2}{xz}\right)  f_{1q}(x) D_{1g}(z) + \frac{T_F}{2} P_{qg}\left(\frac{\xi_2}{xz}\right) f_{1g}(x) D_{1\bar q}(z) \Biggr]~,\label{eq:Matching-u1} \\
l_{1q}^L(x_B,\xi_2,P_{2\perp}) &= \frac{\alpha_s}{2\pi^2 \xi_2 \boldsymbol{P}_{2\perp}^2} \int_{x_B+\xi_2}^1 \frac{dx}{x}z  \Biggl[ \tilde P_{gq}\left(\frac{\xi_2}{xz}\right) f_{1q}(x) G_{1Lg}(z) + \frac{T_F}{2} \tilde P_{qg}\left(\frac{\xi_2}{xz}\right) f_{1g}(x) G_{1L\bar q}(z)\Biggr]~,\label{eq:Matching-l1L}
\end{align}
where $z = \xi_2/(x-x_B)$ and $T_F = 1/2$.
In the convolution, $f_{1q}(x)$ is the quark distribution function, and $D_{1g,\bar q}(z)$ or $G_{1Lg,\bar q}(z)$ are the unpolarized or polarized parton FFs.
The parton splitting functions appeared in the matching coefficients are defined as
\begin{align}
& P_{gq}(z) = C_F \frac{1+(1-z)^2}{z}~, \nonumber \\
& P_{qg}(z) = z^2 + (1-z)^2~, \nonumber \\
& \tilde P_{gq}(z) = C_F(z-2)~, \nonumber \\
& \tilde P_{qg}(z) = -z^2 - (1-z)^2 = -P_{qg}(z)~, \label{eq:Pij}
\end{align}
where $C_F = 4/3$. The matching of antiquark fracture functions can be obtained by charge conjugation. By substituting the matching relations of Eqs.~(\ref{eq:Matching-u1}) and~(\ref{eq:Matching-l1L}) into Eq.~(\ref{eq:CLL-in-FrF}), one obtains the dihadron helicity correlation in the large-$P_{2\perp}$ region. Although both fracture functions scale as $1/\boldsymbol{P}_{2\perp}^2$, the explicit $P_{2\perp}$-dependence cancels in the ratio defining ${\cal C}_{LL}$ in Eq.~(\ref{eq:CLL-in-FrF}).

We now perform a numerical prediction for ${\cal C}_{LL}$ using proton PDFs and $\Lambda$ hyperon FFs as inputs.
For the proton PDFs, we adopt the CTEQ18 set~\cite{Hou:2019efy}.
For the $\Lambda$ hyperon FFs, we employ the DSV parameterization~\cite{deFlorian:1997zj} for both the unpolarized FFs $D_{1q}$ and polarized FFs $G_{1Lq}$ consistently.
Since the DSV parametrization does not distinguish $\Lambda$ from $\bar \Lambda$ for the unpolarized FFs, we take the following assumptions for an estimation~\cite{Chen:2021hdn,Chen:2021zrr}:
\begin{align}
& D_{1q}^{\Lambda}(z) = D_{1\bar q}^{\bar\Lambda}(z) = \frac{1+z}{2}D_{1q}^{\Lambda\bar\Lambda}(z)~, \\
& D_{1\bar q}^{\Lambda}(z) = D_{1q}^{\bar\Lambda}(z) = \frac{1-z}{2}D_{1q}^{\Lambda\bar\Lambda}(z)~.
\end{align}
For the polarized FFs $G_{1Lq}$, the DSV parameterization provides separate descriptions for $\Lambda$ and $\bar{\Lambda}$, with three distinct scenarios on the flavor structures. In scenario-1, only the strange quark contributes to $\Lambda$ polarization at the initial scale, as expected in non-relativistic naive quark model.
In scenario-2, the $u$ and $d$ quarks have negative contributions to the $\Lambda$ polarization, following the estimates of Refs.~\cite{Burkardt:1993zh,Jaffe:1996wp}.
In scenario-3, the $u$, $d$, and $s$ quarks are assumed to contribute equally to the polarized FFs, which is an extreme case considered to explore the possibility that the $u$ and $d$ quarks may also generate $\Lambda$ polarization.

\begin{figure}[!htb]
\centering
\includegraphics[width=0.245\textwidth]{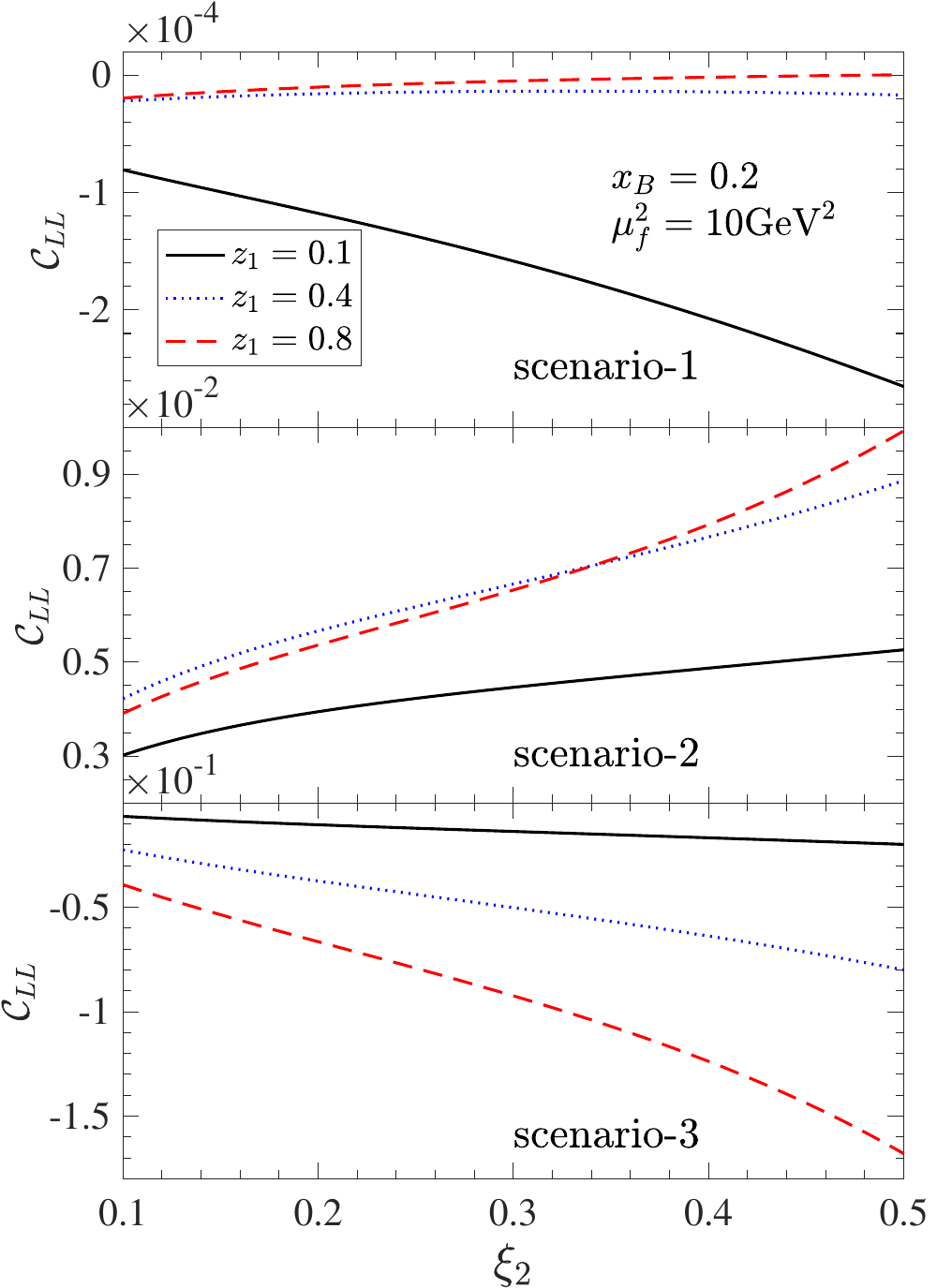}
\includegraphics[width=0.245\textwidth]{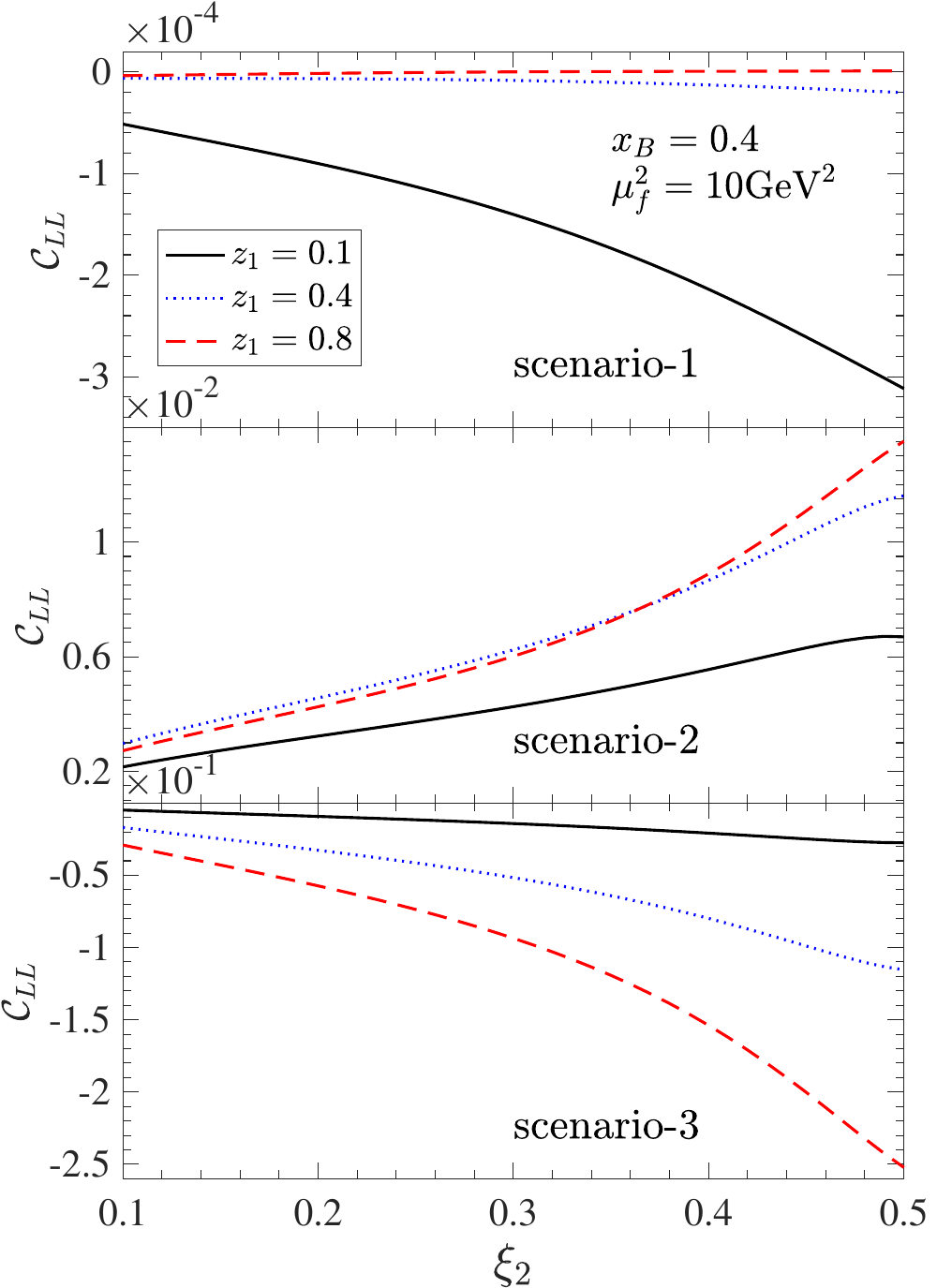}
\includegraphics[width=0.245\textwidth]{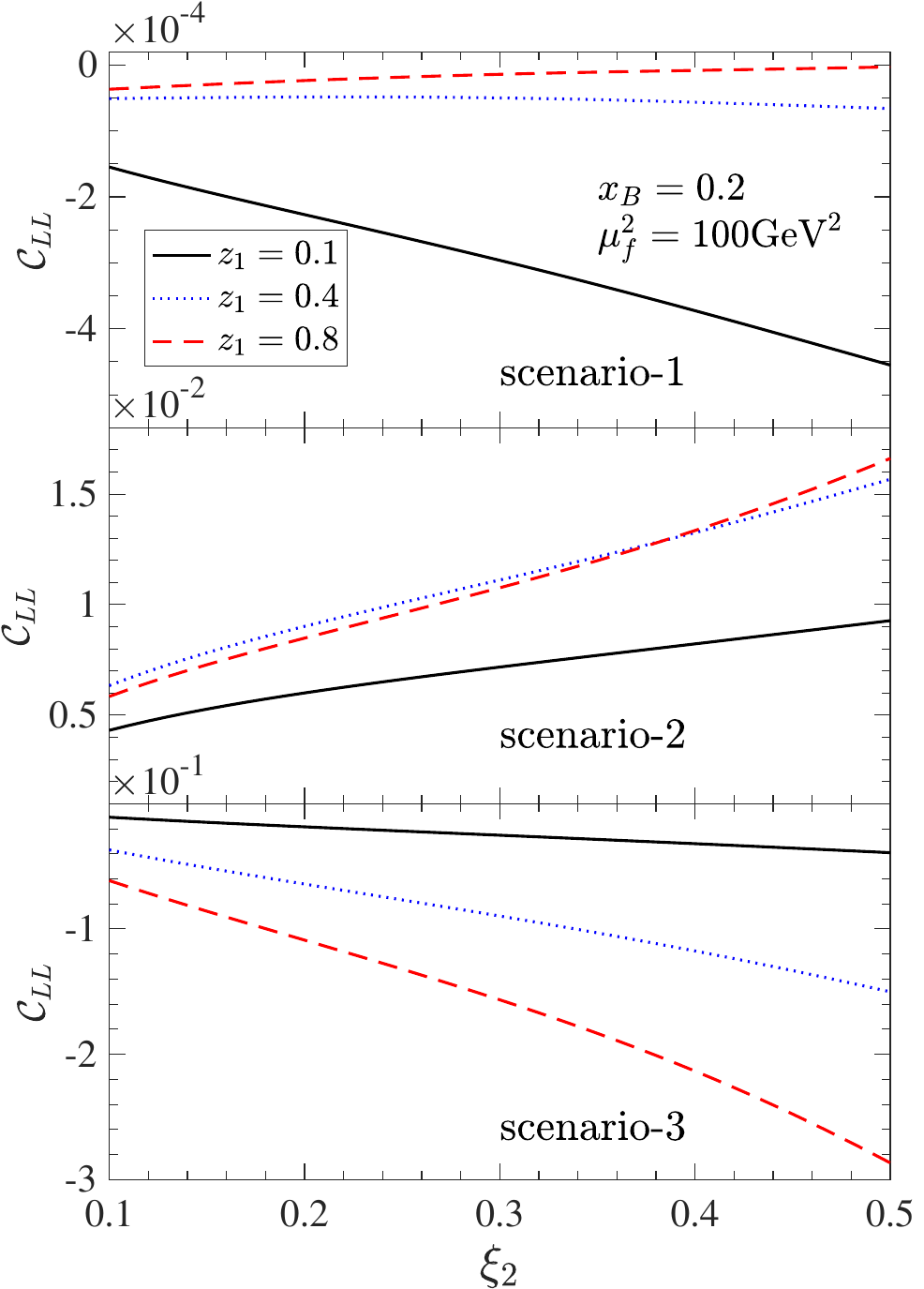}
\includegraphics[width=0.245\textwidth]{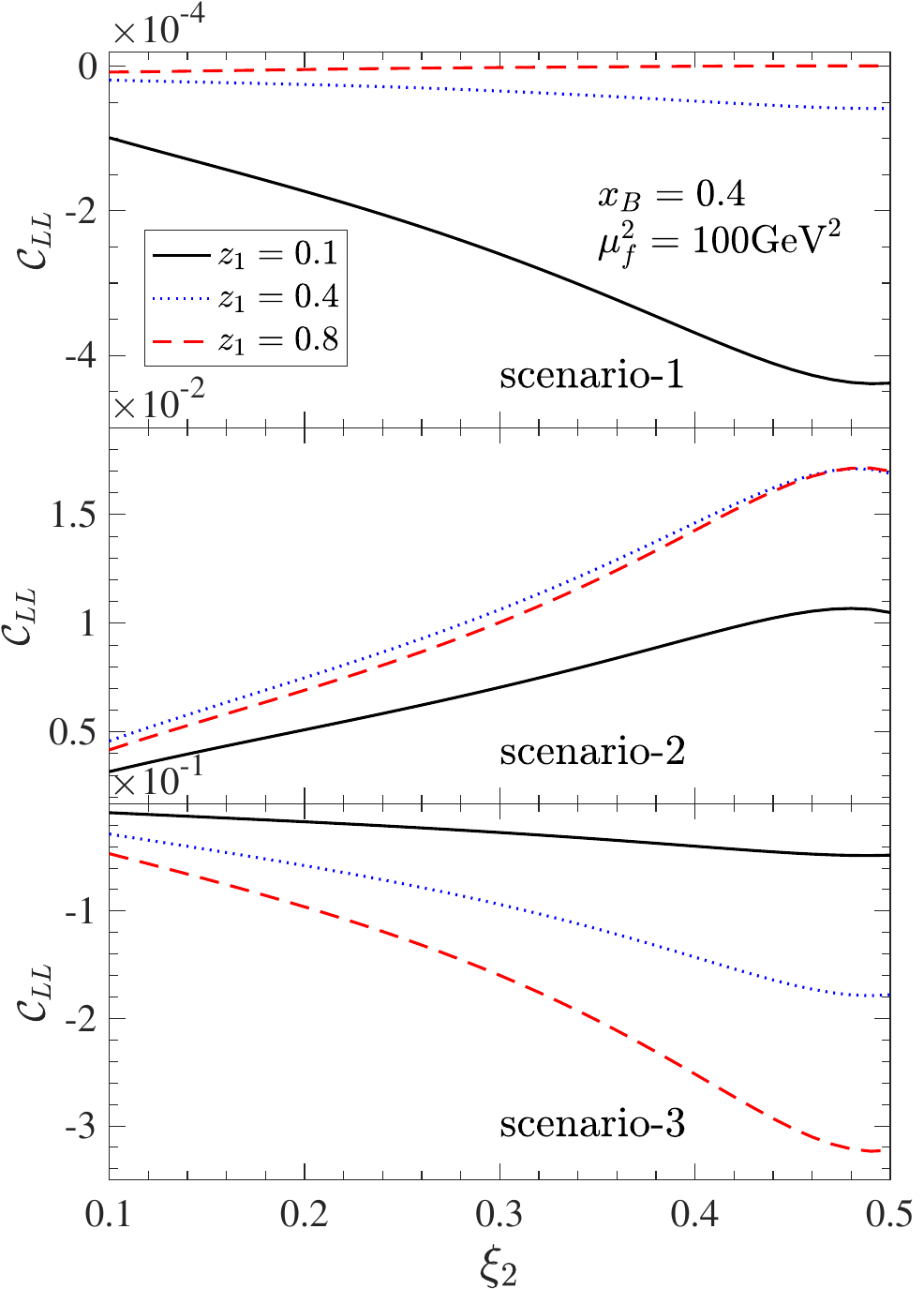}
\caption{Numerical estimations for helicity correlation of $\Lambda$ hyperons in the CFR and TFR as a function of $\xi_2$ for different values of $x_B$ and $z_1$. The typical values of factorization scales are chosen to $\mu_f^2 = 10{\rm GeV}^2$ or $100{\rm GeV}^2$ for experiments at low or high energy scales. Results are shown for three DSV scenarios of polarized $\Lambda$ FFs.}
\label{fig:CLL}
\end{figure}

In the numerical calculation of ${\cal C}_{LL}$, since it depends on three kinematic variables, $x_B$, $\xi_2$ and $z_1$, we show the helicity correlation as a function of $\xi_2$ in Fig.~\ref{fig:CLL} upon choosing several typical values of $x_B$ and $z_1$.
Furthermore, we note that the energy scales of the SIDIS process differ significantly across various experimental setups, necessitating the selection of different factorization scales. Here, as an estimation, we set the factorization scale for the PDFs and FFs to $\mu_f^2 = Q^2$.
To illustrate the energy dependence, we choose a typical value of $\mu_f^2 = 10{\rm GeV}^2$ for the case of relatively low energy scale experiments, such as CLAS12 at the JLab~\cite{Burkert:2020akg}, while for experiments with higher energy scale, such as the Electron-Ion Collider~(EIC)~\cite{Accardi:2012qut}, we choose a typical value of $\mu_f^2 = 100{\rm GeV}^2$.
The results for this two scales are shown in the left and right two subplots of Fig.~\ref{fig:CLL}, respectively.
We also show the results for three DSV scenarios of polarized $\Lambda$ FFs in each subplot.

The results in Fig.~\ref{fig:CLL} show the following distinct features.
First, the helicity correlations differ by orders of magnitude among the three scenarios.
The correlation is suppressed to below $10^{-3}$ for the case of scenario-1. This is because only $s$ quark has dominant contributions for $\Lambda$ polarized FFs in scenario-1, while the $s$ quark parton distribution is small compared to $u$ and $d$ parton distribution. A significant correlation up to 30\% is predicted in scenario-3.
This arises from the extreme assumption of large contribution of $\Lambda$ polarized FFs from $u$ and $d$, together with the large parton distribution of $u$ and $d$ in the proton. The correlation in the case of scenario-2 is somewhere in between.
Second, the helicity correlation in the case of scenario-2 is positive while it is negative for the other two scenarios. This is caused by the assumption of negative contributions to the $\Lambda$ polarization from the $u$ and $d$ quarks in scenario-2. This shows that the helicity correlation is highly sensitive to the flavor separation of $\Lambda$ FFs and fracture functions.

Furthermore, the helicity correlation exhibits a strong dependence on the momentum fraction $\xi_2$ of the $\Lambda$ hyperon in the TFR. This dependence is particularly pronounced when the CFR hadron lies in the small-$z_1$ region of scenario-1, or in the large-$z_1$ region of scenarios-2 and -3.
Interestingly, in scenarios-3, the helicity correlation becomes stronger as the rapidity separation between the CFR and TFR hadrons increases, i.e., the magnitude of ${\cal C}_{LL}$ increases with $z_1$ and $\xi_2$. Scenarios-2 also exhibits this pattern but with less dependence on larger-$z_1$.
By contrast, in scenario-1, the correlation tends to be enhanced at large $\xi_2$ and small $z_1$, i.e., when both hadrons are located closer to the proton direction.

Finally, regarding the energy scale dependence, the results indicate a relatively modest yet non-negligible influence on the helicity correlation. Results at high energies are generally slightly larger than those at low energies. This is primarily attributed to the scale dependence of the gluon polarized fragmentation function in the DSV parameterization, since the dominant process involves quark splitting to a gluon, which subsequently fragments. We note that the gluon polarized fragmentation function is set to zero at the initial scale and evolves to non-zero values at higher scales, the helicity correlation consequently increases with the energy scale. 
These indicate that the helicity correlation can be investigated in SIDIS experiments operating at different energy scales, with a potential advantage at relatively higher energies, such as at the EIC, possibly enhancing its measurability.

The numerical results demonstrate that the dihadron helicity correlation can serve as a clear way to distinguish different polarized FFs scenarios on the flavor contributions.
In particular, while the predicted correlation in scenario-1 is highly suppressed, scenarios-2 and -3 yield sizable values, reaching up to a few percent or even higher in certain kinematic regions. This striking contrast implies that future high-statistics experiments, such as the EIC, with precision on the order of a few percent will be capable of discriminating between these scenarios.
Such differentiation is crucial for understanding the origin of $\Lambda$ polarization and validating theoretical models of spin transfer mechanisms in both the current and target fragmentation regions.

\section{Summary}
\label{sec:summary}

We have studied helicity correlation of two $\Lambda$ hyperons produced in unpolarized SIDIS, with one in the CFR and the other in the TFR. The correlation is determined by the polarized fragmentation functions and the polarized fracture functions.
In particular, we consider the TFR hadron in the intermediate transverse momentum region $\Lambda_{\text{QCD}} \ll P_{2\perp} \ll Q$ and perform a perturbative matching of the unpolarized and polarized quark fracture functions onto the proton PDFs and $\Lambda$ FFs. Employing the DSV parametrization of $\Lambda$ FFs, we evaluate this dihadron helicity correlation under three scenarios characterizing the flavor dependence of the polarized FFs. The results show a strong sensitivity to these scenarios. Moreover, a pronounced dependence on the TFR momentum fraction $\xi_2$ is predicted, varying with the rapidity separation between the CFR and TFR hadrons. Interestingly, in scenarios-2 and -3, the helicity correlation becomes stronger as the rapidity separation of the dihadron increases, while in scenario-1 it is enhanced when both hadrons are detected closer to the proton direction.
The results also show a modest scale dependence, where correlation increases with energy due to the evolution of the gluon polarized fragmentation function.

These findings demonstrate that the dihadron helicity correlation serves as a sensitive observable for probing the flavor structure of polarized fragmentation processes in both the CFR and TFR, thereby providing new constraints on spin-dependent fragmentation and fracture functions. Future SIDIS measurements, particularly at the EIC, will be essential for disentangling the flavor dependence of spin transfer and for establishing a more comprehensive understanding of spin-dependent hadronization dynamics.

\section*{Acknowledgments}
We thank Jian-Ping Ma for useful discussions. This work is supported by Natural Science Foundation of China under grant No.~12405156, 12005122, the Shandong Province Natural Science Foundation under grant No.~2023HWYQ-011, ZFJH202303, ZR2020QA082, the Taishan fellowship of Shandong Province for junior scientists, and Youth Innovation Team Program of Higher Education Institutions in Shandong Province (Grant No. 2023KJ126) as well as the Research Council of Finland, the Centre of Excellence in Quark Matter and projects 338263 and 359902, and the European Research Council (ERC, grant agreements No. ERC-2023-101123801 GlueSatLight and No. ERC-2018-ADG-835105 YoctoLHC). The content of this article does not reflect the official opinion of the European Union and responsibility for the information and views expressed therein lies entirely with the authors.

\appendix
\section{The matching of the fracture functions}
\label{sec:app}
In the region $\Lambda_{\text{QCD}} \ll P_{2\perp} \ll Q$, the matching of fracture functions onto PDFs and FFs is conveniently performed in the light-cone gauge~\cite{Chen:2021vby}.
In this gauge, only two diagrams shown in Fig.~\ref{fig:Matching} will contribute.
The contribution of Fig.~\ref{fig:Matching}(a) to the matrix element of Eq.~(\ref{eq:MFrF}) is calculated as
\begin{align}
\hat {\cal M}_{q,ij}\Big\vert_{\ref{fig:Matching}(a)} =& \frac{1}{(2\pi)^3 2\xi_2} \int dx \frac{dz}{z^2} P^+  \delta(xP^+ - x_BP^+ - k_g^+) \Biggl[ \frac{i\gamma\cdot (xP-k_g)}{(xP-k_g)^2} (-ig_s \gamma_\mu T^a) \nonumber\\
& \qquad\qquad \times \frac{1}{2N_c} f_{1q}(x) \gamma^- (ig_s \gamma_\nu T^a) \frac{-i\gamma\cdot (xP-k_g)}{(xP-k_g)^2} \Biggr]_{ij} \Bigl[ \frac{1}{2}N^{\mu\nu}(k_g) D_{1g}(z) - \frac{i}{2}E^{\mu\nu}(k_g) \lambda_2 G_{1Lg}(z) \Bigr], \label{eq:matching-quark}
\end{align}
where $k_g^\mu = P_2^\mu/z$ denotes the momentum of the fragmenting gluon.
The Lorentz tensors $N^{\mu\nu}(k_g)$ and $E^{\mu\nu}(k_g)$ are defined as $N^{\mu\nu}(k_g) \equiv - g^{\mu\nu} + (n^\mu k_g^\nu + n^\nu k_g^\mu)/n\cdot k_g$, and $E^{\mu\nu}(k_g) \equiv k_{g\alpha} n_\beta \varepsilon^{\mu\nu \alpha\beta}/n\cdot k_g$ with $\varepsilon^{0123}=1$ and the light-cone vector $n^\mu = (0,1,\boldsymbol{0}_\perp)$.
They satisfy $N^{\mu\nu}(k_g) N_{\mu\nu}(k_g) = E^{\mu\nu}(k_g) E_{\mu\nu}(k_g) = 2$ and $N^{\mu\nu}(k_g) E_{\mu\nu}(k_g) = 0$.
Here, $f_{1q}(x)$ is the standard quark collinear distribution function, $D_{1g}(z)$ and $G_{1Lg}(z)$ are the unpolarized and polarized gluon collinear FFs.
They come from the decomposition of the density matrices represented by the gray boxes in Fig.~\ref{fig:Matching}, e.g.,
\begin{align}
\int \frac{d\eta^-}{2\pi} e^{i\eta^- xP^+} \langle P|\bar \psi_j(0) \psi_i(\eta^-) |P\rangle = \frac{1}{2N_c} \left[ f_{1q}(x) \gamma^- \right]_{ij} + \cdots.
\end{align}
Similarly, the contribution of Fig.~\ref{fig:Matching}(b) is calculated as
\begin{align}
\hat {\cal M}_{q,ij}\Big\vert_{\ref{fig:Matching}(b)} =& \frac{1}{(2\pi)^3 2\xi_2} \int \frac{dx dz}{xP^+ z^2} P^+ \delta(xP^+ - x_BP^+ - k_g^+) \Biggl[ \frac{i\gamma\cdot (xP-k_{\bar q})}{(xP-k_{\bar q})^2} (-ig_s\gamma_\mu T^a) \nonumber\\
&\qquad\qquad \times \bigl( \gamma\cdot k_{\bar q} D_{1\bar q}(z) - \gamma_5\gamma\cdot k_{\bar q} \lambda_2 G_{1L\bar q}(z)\bigr)  (ig_s\gamma_\nu T^a) \frac{-i\gamma\cdot (xP-k_{\bar q})}{(xP-k_{\bar q})^2} \Biggr]_{ij} \Bigl[ -\frac{1}{2}g_\perp^{\mu\nu} f_{1g}(x) \Bigr],\label{eq:matching-gluon}
\end{align}
where $k_{\bar q}^\mu = P_2^\mu/z$ denotes the momentum of the fragmenting antiquark. The transverse metric is defined as $g_\perp^{\mu\nu} \equiv g^{\mu\nu} - \bar n^\mu n^\nu - \bar n^\nu n^\mu$ with the light-cone vector $\bar n^\mu = (1,0,\boldsymbol{0}_\perp)$.
$D_{1\bar q}(z)$ or $G_{1L\bar q}(z)$ are the unpolarized or polarized FFs for the antiquark, $f_{1g}(x)$ is the unpolarized gluon distribution function.

The total contribution to the matrix element is given by
\begin{align}
\hat {\cal M}_{q,ij} = \hat {\cal M}_{q,ij}\Big\vert_{\ref{fig:Matching}(a)} + \hat {\cal M}_{q,ij}\Big\vert_{\ref{fig:Matching}(b)}.
\end{align}
From the decomposition of the matrix element in terms of the fracture functions given in Eq.~(\ref{eq:DecFrF}), we have
\begin{align}
u_{1q} &= {\rm Tr} \left[ \frac{\gamma^+}{2} \hat {\cal M}_{q} \right], \qquad \lambda_2 l_{1q}^L = {\rm Tr} \left[ \frac{\gamma^+\gamma_5}{2} \hat {\cal M}_{q} \right].
\end{align}
Working out the trace, we obtain
\begin{align}
u_{1q}(x_B,\xi_2,P_{2\perp}) &= \frac{\alpha_s}{2\pi^2\xi_2 \boldsymbol{P}_{2\perp}^2} \int_{\frac{\xi_2}{1-x_B}}^1 \frac{dz}{z^2} \Bigl[ C_F z^2(x_B^2/x^2+1) f_{1q}(x) D_{1g}(z) + \frac{T_F}{2} \frac{\xi_2}{zx^3}(z^2x_B^2+\xi_2^2) f_{1g}(x) D_{1\bar q}(z) \Bigr], \label{eq:matching-u1-raw} \\
l_{1q}^L(x_B,\xi_2,P_{2\perp}) &= \frac{\alpha_s}{2\pi^2\xi_2 \boldsymbol{P}_{2\perp}^2} \int_{\frac{\xi_2}{1-x_B}}^1 \frac{dz}{z^2} \Bigl[ C_F \frac{\xi_2}{x^2}(\xi_2-2xz) f_{1q}(x) G_{1Lg}(z) - \frac{T_F}{2} \frac{\xi_2}{zx^3}(z^2x_B^2+\xi_2^2) f_{1g}(x) G_{1L \bar q}(z)\Bigr], \label{eq:matching-l1L-raw}
\end{align}
where $C_F=4/3$, $T_F = 1/2$, and $x = x_B+\xi_2/z$.
The limits of integration for $z$ are determined by the conditions of $0<x=x_B+\xi_2/z<1$ in the PDFs and $0<z<1$ in the FFs.
By making the substitution $z\to \xi_2/(x-x_B)$, we can convert the integration variable from $z$ to $x$ and, after simplification, rewrite Eqs.~(\ref{eq:matching-u1-raw}) and (\ref{eq:matching-l1L-raw}) in the form shown in Eqs.~(\ref{eq:Matching-u1})-(\ref{eq:Pij}).

\end{document}